# High-Speed Gate Driver Using GaN HEMTs for 20-MHz Hard Switching of SiC MOSFETs

Takafumi Okuda and Takashi Hikihara

Department of Electrical Engineering, Kyoto University, Katsura, Nishikyo, Kyoto 615-8510

Email address: t-okuda@dove.kuee.kyoto-u.ac.jp

In this paper, we investigated a gate driver using a GaN HEMT push-pull configuration for the high-frequency hard switching of a SiC power MOSFET. Low on-resistance and low input capacitance of GaN HEMTs are suitable for a high-frequency gate driver from the logic level, and robustness of SiC MOSFET with high avalanche capability is suitable for a valve transistor in power converters. Our proposed gate driver consists of digital isolators, complementary Si MOSFETs, and GaN HEMTs. The GaN HEMT push-pull stage has a high driving capability owing to its superior switching characteristics, and complementary Si MOSFETs can enhance the control signal from the digital isolator. We investigated limiting factors of the switching frequency of the proposed gate driver by focusing on each circuit component and proposed an improved driving configuration for the gate driver. As a result, 20-MHz hard switching of a SiC MOSFET was achieved using the improved gate driver with GaN HEMTs.

## I. INTRODUCTION

Widegap semiconductor materials such as silicon carbide (SiC) and gallium nitride (GaN) have high critical electric field strength, which is attractive for high-voltage power devices [1-3]. In SiC, high-quality bulk crystals with a large diameter (~6 inch) can be obtained with sublimation method, and SiC homoepitaxial layers can be grown on SiC bulk crystals with chemical vapor deposition (CVD) [4,5]. Thus, a vertical structure is typically utilized for SiC power devices. SiC metal-oxide-semiconductor field-effect transistors (MOSFETs) with a high blocking voltage (> 1 kV) and low on-resistance (< 100 m$\Omega$cm$^2$) have been reported [6-9]. SiC MOSFETs also exhibit a sufficient avalanche capability [10,11], which is an advantage as a valve transistor for high-voltage and high-current power converters.

In GaN, even though no mass production of vertical power devices was achieved due to a lack of high-quality and large GaN bulk crystals, the heteroepitaxial growth of the mixed alloy in the III-nitrides is available on a Si substrate [12,13]. Two-dimensional electron gas (2DEG) is generated at a heterojunction of AlGaN/GaN, and GaN-based high electron mobility transistors (HEMTs) have been developed [14-18]. Although GaN HEMTs still have poor avalanche capability, their on-resistance and input capacitance are lower than those of MOS-based devices, which enable high-frequency operation with low power consumption.

Since SiC MOSFETs are suitable for valve transistors in power converters and GaN HEMTs are applicable to gate drivers with high driving capability, we previously proposed a high-speed gate driver (Prototype-A) based on a GaN-HEMT push-pull configuration for driving a SiC MOSFET [19]. By using the proposed gate driver, 10-MHz hard switching of SiC MOSFETs was obtained. Although there are other reports on gate drivers using GaN HEMTs [20-23], the driving target is GaN-HEMT valve switches or soft-switching circuits such as resonant converters. In this study, on the other hand, we focus on the high-frequency hard switching of SiC MOSFETs because power converters with hard switching are suitable for handling various kinds of loads. We enhance the switching frequency up to ISM radio bands (Type-B worldwide: 13.56 or 27.12 MHz) that are reserved internationally for radio frequency energy. A high-speed gate driver with a high driving capability is necessary to drive SiC MOSFETs at high switching frequency.

In this study, we investigated the limiting factors of the switching frequency in the proposed gate driver and achieve 20-MHz hard switching of SiC MOSFETs with an improved gate driver. The 20-MHz frequency (20.34 MHz) corresponds to the third harmonic of 6.87 MHz that is also reserved for Type-A ISM radio bands.

## II. CONFIGURATION OF GATE DRIVER

A schematic of our fabricated gate driver is shown in Fig. 1. The GaN-HEMT push-pull configuration is employed for the driving stage. The dual control signals are generated with a function generator (Tektronix, AFG3102C) and isolated by digital isolators (Silicon Labs, Si8610). The digital isolator, which consists of an RF transmitter and receiver for isolation, is commonly used for digital signal transmission to isolate the transmission line. The output signals from the digital isolators are enhanced through complementary Si MOSFETs (ROHM, US6M1) and finally connected to the GaN-HEMT push-pull stage. The GaN HEMTs are distributed from ROHM for Prototype-A and EPC (EPC2014C) for Prototype-B, both of which are fabricated as normally-off (enhancement-mode) transistors.

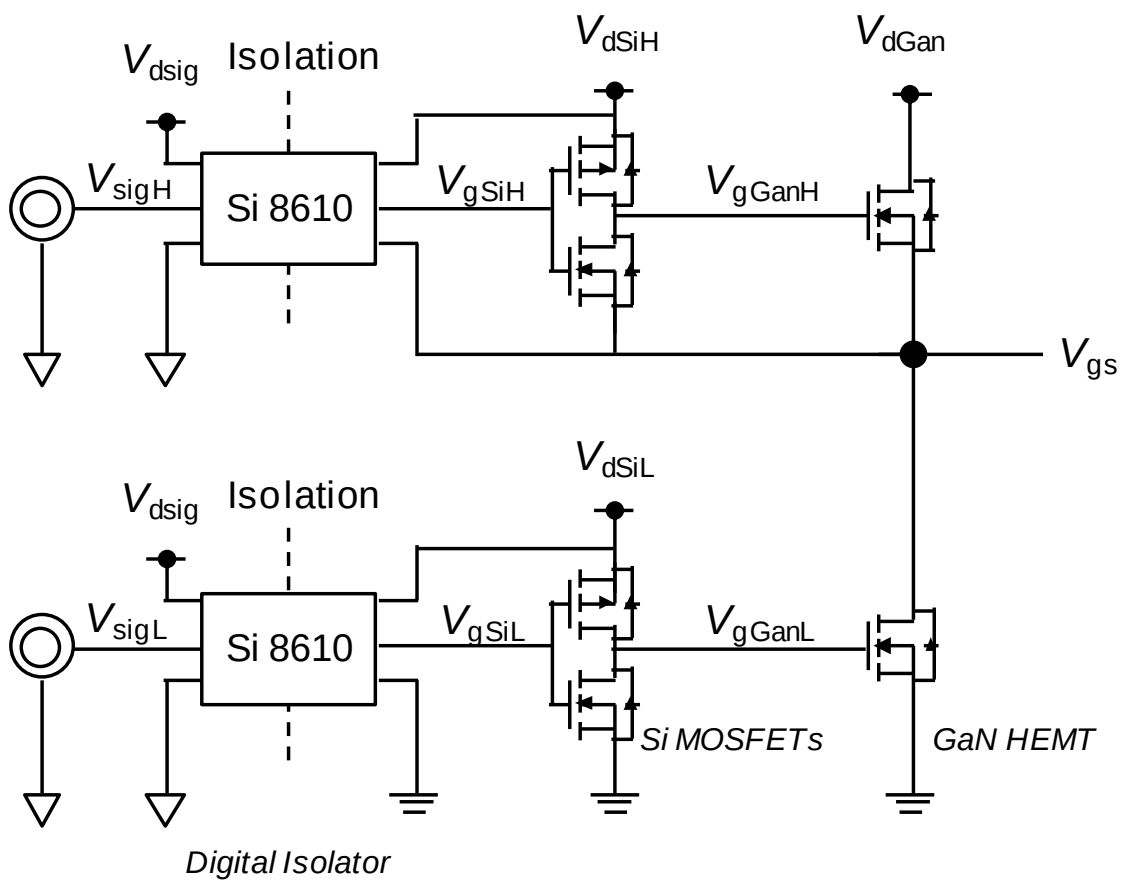

Fig. 1 Schematic of fabricated gate driver using GaN HEMT push-pull configuration.

Four isolated floating voltage sources (Matsusada, P4K36-1) were used for the proposed gate driver. Input power supply $V_{dsig}$ for the digital isolators was set at 5 V, and an output power supply for the digital isolators at the high and low sides ($V_{dSiH}$ and $V_{dSiL}$) was employed. $V_{dSiH}$ and $V_{dSiL}$ also supply power to the complementary Si MOSFETs at the high and low sides. Here $V_{dSiH}$ and $V_{dSiL}$ were varied from 3 to 5 V to investigate the switching characteristics of the complementary Si MOSFETs. $V_{dGan}$, which supplies a driving voltage at the GaN-HEMT push-pull configuration, is fixed at 18 V to drive the SiC MOSFETs.

Note that a bootstrap capacitor is not employed for $V_{dSiH}$ in this study. According to our previous study [19], a high-side voltage source $V_{dSiH}$ was generated by a bootstrap capacitor from the low-side voltage source $V_{dSiL}$, which simplifies the voltage supply at the high side. However, the forward voltage drop of the bootstrap diode is 0.8-1.0 V, and the high-side supply voltage is lower than the low-side. As described below, the supply voltage for the complementary Si MOSFETs strongly affects their switching characteristics. In order to overcome the upper limit of the driving frequency in our proposed gate driver, we removed bootstrap capacitor for the high-side voltage supply and used another isolated floating voltage supply $V_{dSiL}$ to balance the supply voltage of the high-side and low-side Si MOSFETs.

The fabricated gate drivers are shown in Fig. 2. Prototype-A, which was fabricated in our previous study [19], has test terminals on the board, but the parasitic inductance increases with extra circuit patterns. In order to reduce the parasitic inductance, in this study we fabricated a small-sized gate driver, Prototype-B. In Section III, Prototype-A is investigated to clarify the limiting factors of the high-frequency operation using their test terminals. In Section IV, we use Prototype-B to achieve 20-MHz hard switching of SiC MOSFETs and measure the switching waveforms by an oscilloscope (Tektronix, MDO4104) with a passive voltage probe (Tektronix, TPP1000) and a current probe (Tektronix, TCP0030).

[Prototype-A] [Prototype-B]

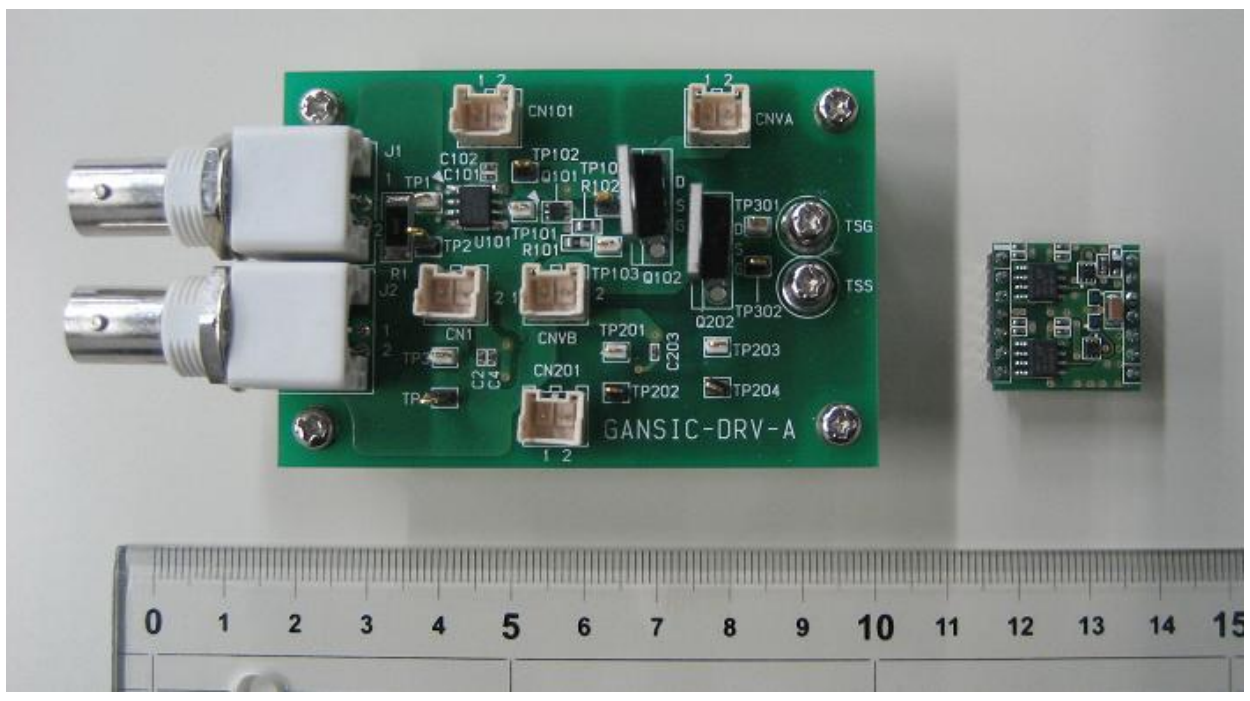

Fig.2 Photograph of fabricated gate drivers: Prototype-A and Prototype-B.

## III. INVESTIGATION OF LIMITING FACTORS OF HIGH-FREQUENCY OPERATION

In this section, we investigate the limiting factors of high-frequency operation in our proposed gate driver at frequencies exceeding 10 MHz by focusing on each circuit component, the digital isolators, the Si MOSFETs, and the GaN HEMTs.

### A. Propagation Characteristics of Digital Isolators

We investigated the propagation characteristics of digital isolators Si8610 and measured the switching behaviors at the low-side digital isolator of the gate driver. The output characteristics of the digital isolator are shown in Fig. 3. The switching frequency was set at 1 or 20 MHz, and supply voltage $V_{\mathrm{dSiL}}$ for the digital isolator was set at 3.5 V. The digital isolator worked at 1 MHz, and a clear output signal $V_{\mathrm{gSiL}}$ was obtained, whereas output signal $V_{\mathrm{gSiL}}$ was strongly degraded at 20 MHz owing to the poor propagation capability of the digital isolator at high frequency. The output voltage of the on-state mode decreased to 2.5 V, and the output voltage of the off-state mode increased to 1.2 V. Although the data rate of the digital isolator is 150 Mbps according to the datasheet, its driving capability of the digital isolator is insuffient to drive the complementary Si MOSFETs at 20 MHz. The propagation delay at the digital isolator is approximately 8 ns according to the datasheet. This value is consistent with the delay obtained by the experimental result shown in Fig. 3.

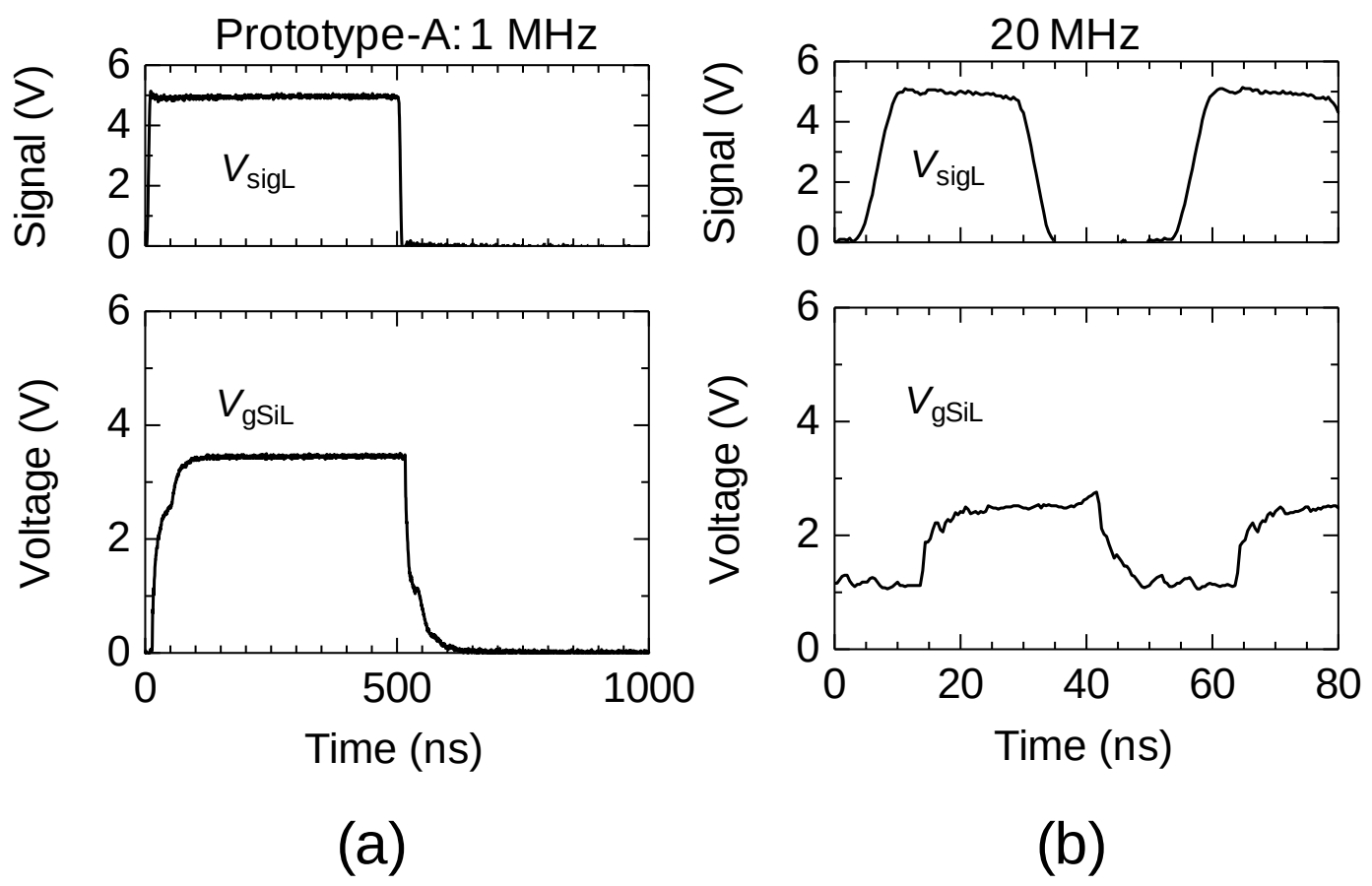

Fig.3 Output characteristics of digital isolator (Si8610): (a) 1 MHz and (b) 20 MHz, measured in proposed gate driver Prototype-A.

Since the output signal from the digital isolator was degraded at high switching frequency, the output signal from the digital isolator must be enhanced. This is why complementary Si MOSFETs are employed in the next driving stage. The output characteristics of the Si MOSFETs at 1 MHz are shown in Fig. 4. Supply voltage $V_{dSiL}$ for the Si MOSFETs varied from 3 to 5 V. Their output voltage exhibited faster switching behaviors than the digital isolator, suggesting that the driving capability is enhanced through complementary Si MOSFETs.

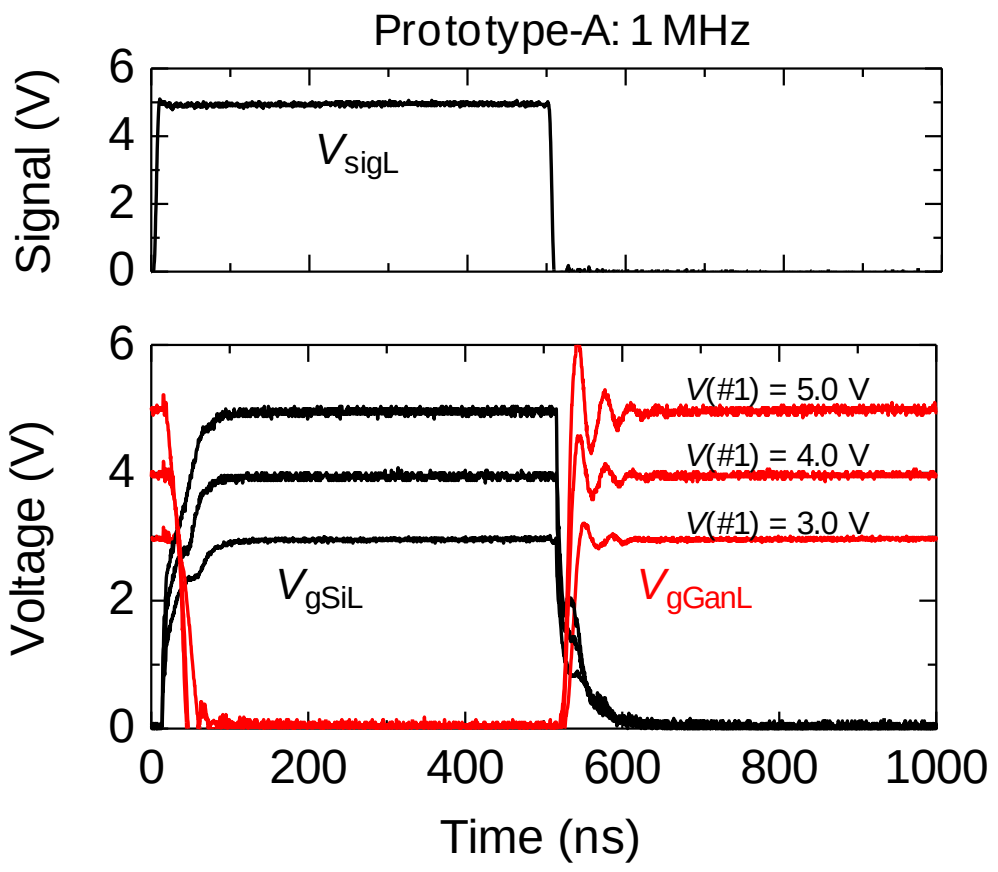

Fig. 4 Output characteristics of digital isolator ($V_{gSiL}$) and Si MOSFETs ($V_{gGanL}$) at frequency of 1 MHz measured in proposed gate driver Prototype-A. Supply voltage ($V_{dSiL}$) varied from 3.0 to 5.0 V.

B. Limitation of Source Voltage for Si MOSFETs

Si MOSFETs can enhance the control signal from the digital isolator, but supply voltage $V_{dSiL}$ for Si MOSFETs is complicated, as described below. A higher supply voltage would enhance the driving capability, while the penetration current through the complementary Si MOSFETs increases with additional supply voltage. Hence, an appropriate supply condition is required to avoid the thermal runaway induced by the penetration current at high switching frequency.

We measured the average supply currents from the voltage source to the Si MOSFETs with various supply voltages $V_{dSiL}$ to investigate the effect of the penetration current at the switching. The supply currents to the Si MOSFETs plotted against the switching frequency are shown in Fig. 4. Supply voltage $V_{dSiL}$ was varied from 3.0 to 4.5 V. The supply current increases with increasing the switching frequency as well as increasing the supply voltage $V_{dSiL}$. The package temperature of the Si MOSFETs is monitored by an infrared camera (optris PI-160), which allows for exact measurements from an object size of 1.5 mm with a measurement speed of 120 Hz. It was found that the package temperature of the Si MOSFETs rose to 100°C at supply currents higher than approximately 200 mA, resulting in a thermal runaway of the Si MOSFETs. In the case of a supply voltage $V_{dSiL}$ of 4.5 V, the supply current reached 200 mA at 10 MHz. In order to operate the Si MOSFET at 20 MHz, we employed a supply voltage $V_{dSiL}$ of 3.8 V. The penetration current in the Si MOSFETs was found to be one of the limiting factors to enhance the switching frequency of the gate driver. For 20-MHz hard switching in the proposed gate driver, a bootstrap diode was removed to balance the supply voltage of the high-side and low-side Si MOSFETs. It is important to choose complementary Si MOSFETs with low power consumption (low penetration current) for high-frequency switching.

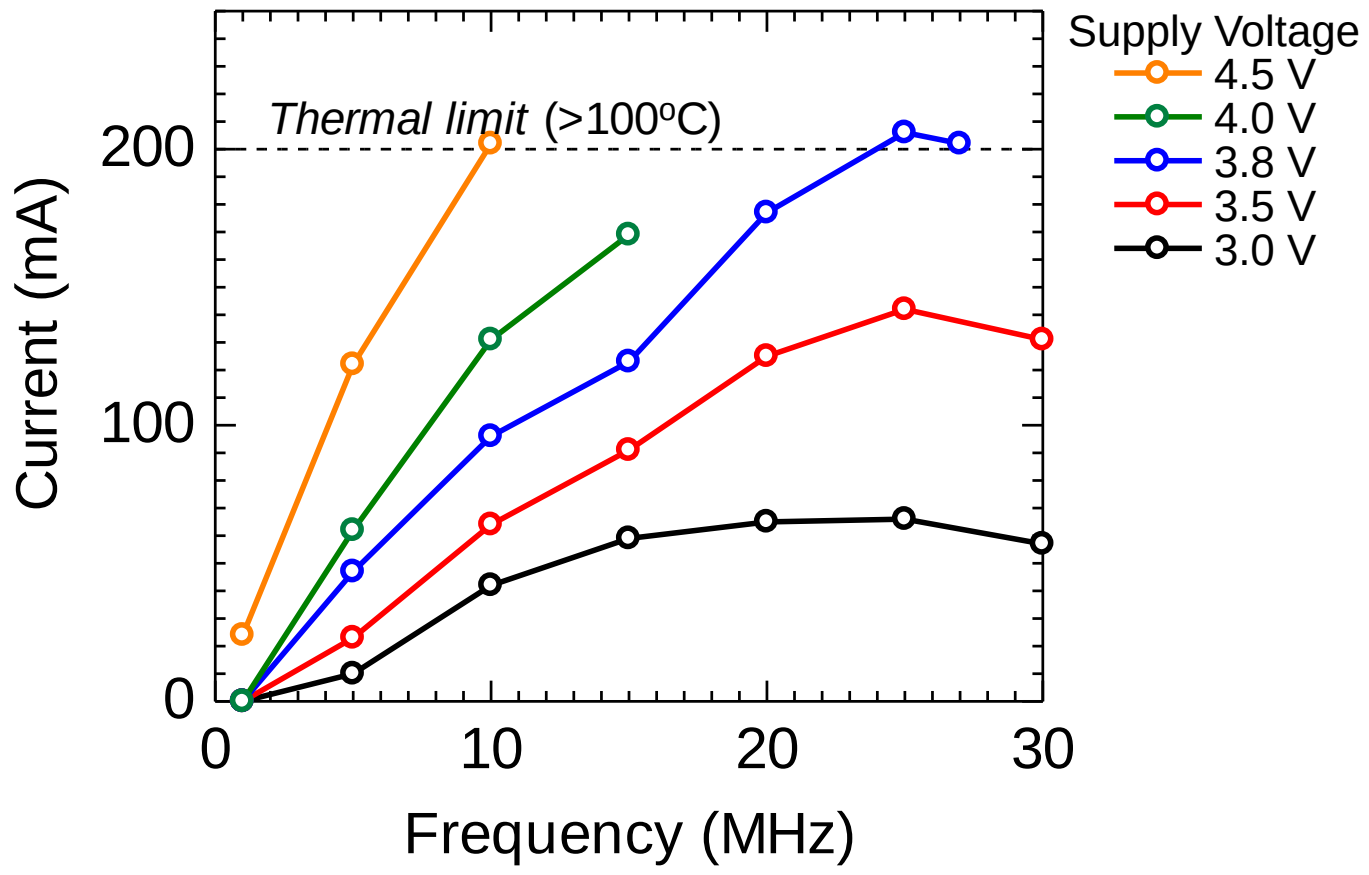

Fig.5 Current consumption of Si MOSFETs plotted against switching frequency. Supply voltage ($V_{dSiL}$) was varied from 3.0 to 4.5 V. Thermal limit is defined at a device temperature over 100°C.

C. Limitation of Driving Capability of Si MOSFETs

Complementary Si MOSFETs can enhance the output signal from the digital isolator, as described above. The output voltage $V_{gSiL}$ from the digital isolator and $V_{gGanL}$ from the Si MOSFETs are shown in Fig. 6. The switching frequency varied from 1 to 30 MHz and the supply voltage $V_{dSiL}$ to the Si MOSFETs was set at 3.5 V. The output signal from the digital isolator was degraded by increasing the switching frequency, but the Si MOSFETs enhanced the output signal. Since threshold voltages of the Si MOSFETs exist between the output signals at the on-state (2.5 V at 20 MHz) and the off-state (1.2 V at 20 MHz) from the digital isolators, the Si MOSFETs narrowly worked at high switching frequency.

However, the maximum value of output voltage $V_{gGanL}$ from the Si MOSFETs decreased to 2.5 V at 30 MHz. The driving capability of the Si MOSFETs is inadequate at 30 MHz.

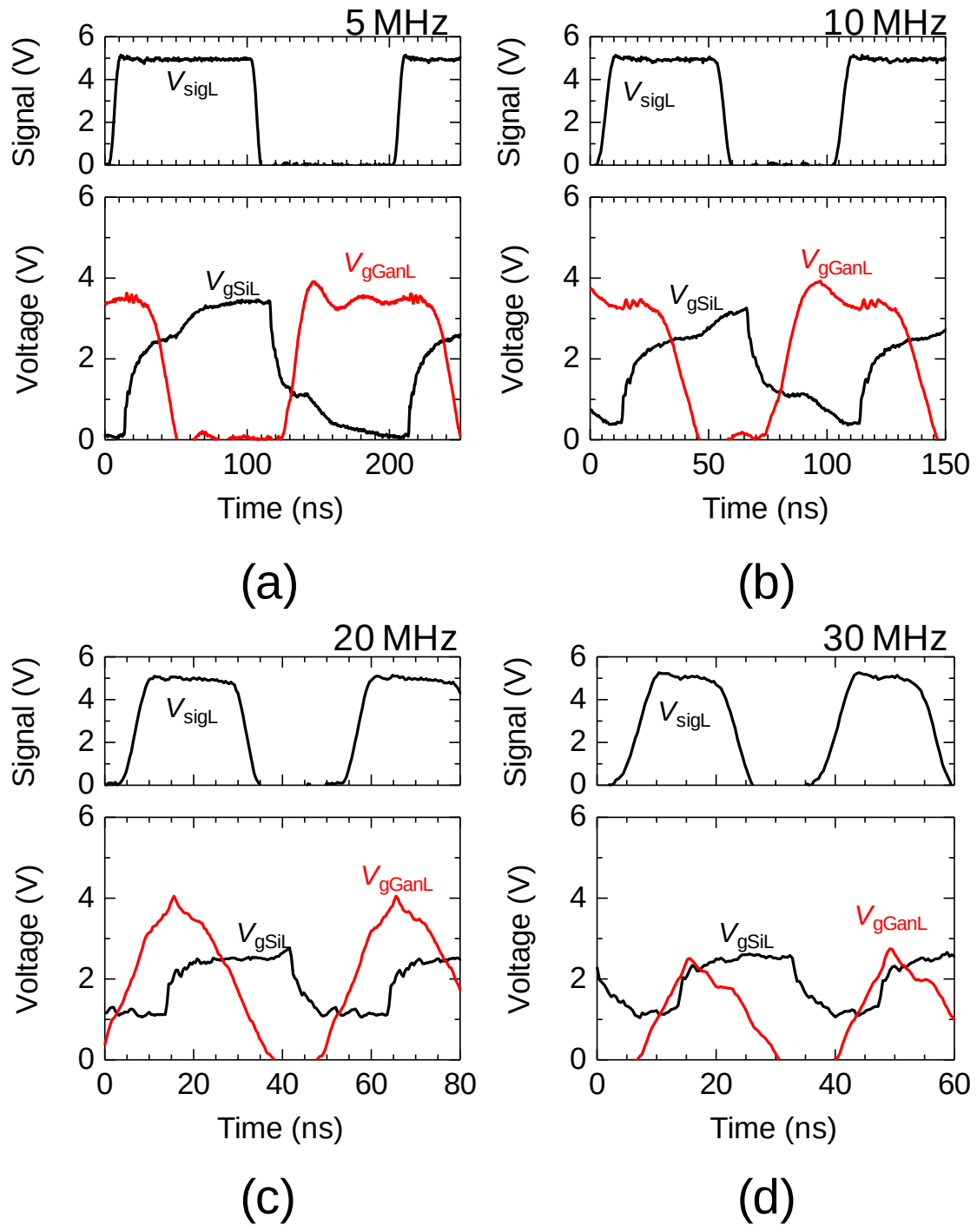

Fig.6 Output characteristics of a digital isolator (Si8610): (a) 1 MHz, (b) 10 MHz, (c) 20 MHz, and (d) 30 MHz, measured in proposed gate driver Prototype-A.

The maximum values of the output voltage $V_{gGanL}$ are plotted against the switching frequency in Fig. 7. The supply voltage $V_{dSiL}$ to the Si MOSFETs was set at 3.5 V. Note that we observed the overshoot of the output voltage, and the maximum value of the output voltage increased to 3.9 V compared to a supply voltage of 3.5 V. The output voltage decreased with increasing the switching frequency over 20 MHz, resulting in 2.5 V at 30 MHz. In order to further enhance the switching frequency, we must employ Si MOSFETs with a higher driving capability or GaN HEMTs with a smaller input capacitance in the gate driver.

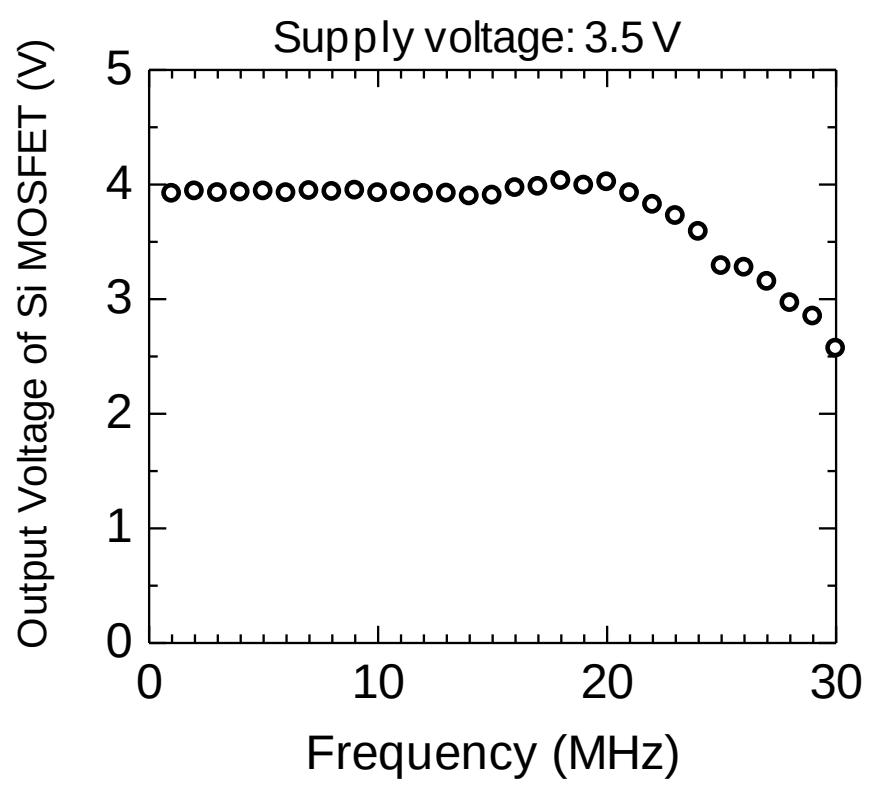

Fig.7 Output voltage ($V_{gGanL}$) of Si MOSFETs plotted against switching frequency. Supply voltage ($V_{dSiL}$) was set at 3.5 V.

D. Output Characteristics of GaN-HEMT Push-Pull Driver

The above investigations were conducted in a gate driver Prototype-A that has testing terminals. Here, we compare the output characteristics of Prototypes-B and Prototype-A.

The output characteristics of the gate drivers of Prototype-A and Prototype-B are shown in Fig. 8. The measurements with the oscilloscope were triggered by input signals from the function generator. The switching frequency was set at 14 MHz, and supply voltages $V_{dSiL}$ and $V_{dSiH}$ were set at 3.8 V and the supply voltage $V_{dGan}$ was set at 18 V. The output of the gate driver was open (no driving target). The voltage surge was significantly more reduced in the output characteristics of Prototype-B than Prototype-A, probably caused by the reduction of the parasitic inductances in the gate driver owing to its downsizing. The gate driver Prototype-B exhibited higher driving capability with a lower surge voltage than Prototype-A.

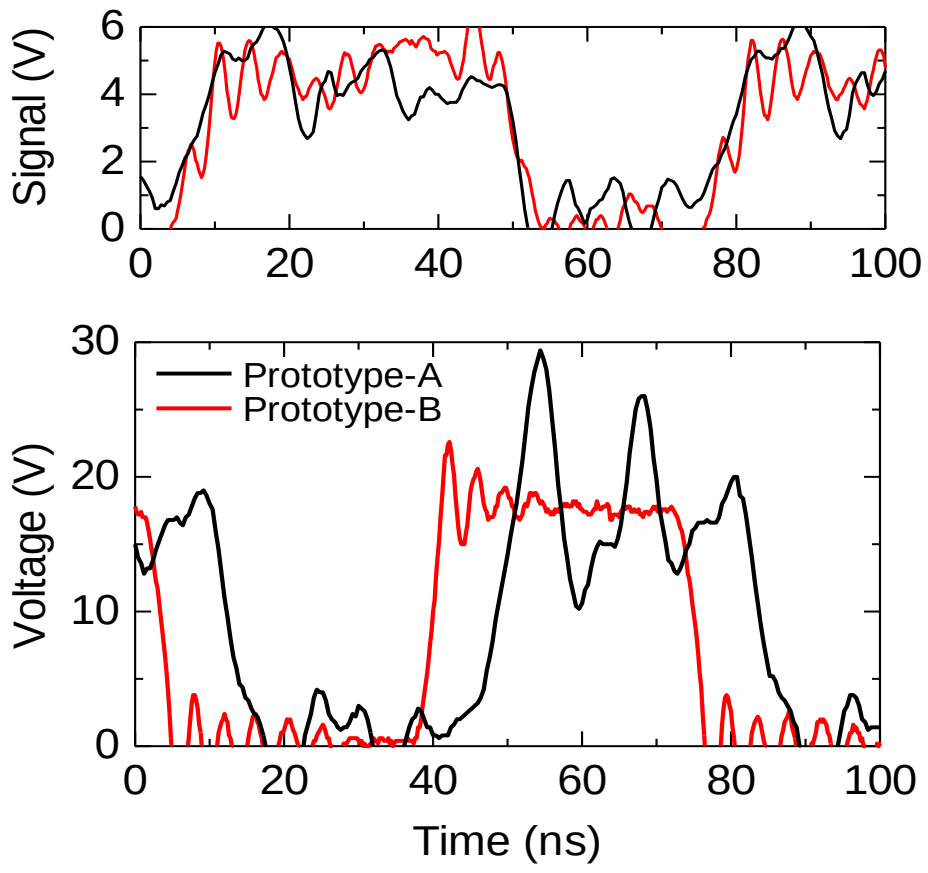

Fig. 8 Output characteristics of GaN HEMTs in the gate drivers of Prototype-A and Prototype-B. Duty ratios of high-side and low-side input signals were set to 0.6.

IV. HIGH-FREQUENCY HARD SWITCHING OF SIC MOSFETS

Next the hard switching of SiC MOSFETs was investigated using Prototype-B. The schematic of the testing circuit for hard switching is shown in Fig. 9. The link voltage was set at 50 V, and the current limiter was a 100-Ω resister. We used a 1200-V 10-A SiC MOSFET (ROHM, SCT2450KE). No gate resistance was employed here, and gate driver Prototype-B was directly connected to the SiC MOSFET to investigate the maximum driving capability of the gate driver. Supply voltage $V_{dsig}$ was set at 5 V, supply voltages $V_{dSiL}$ and $V_{dSiH}$ were set at 3.8 V, and supply voltage $V_{dGan}$ was set at 18 V.

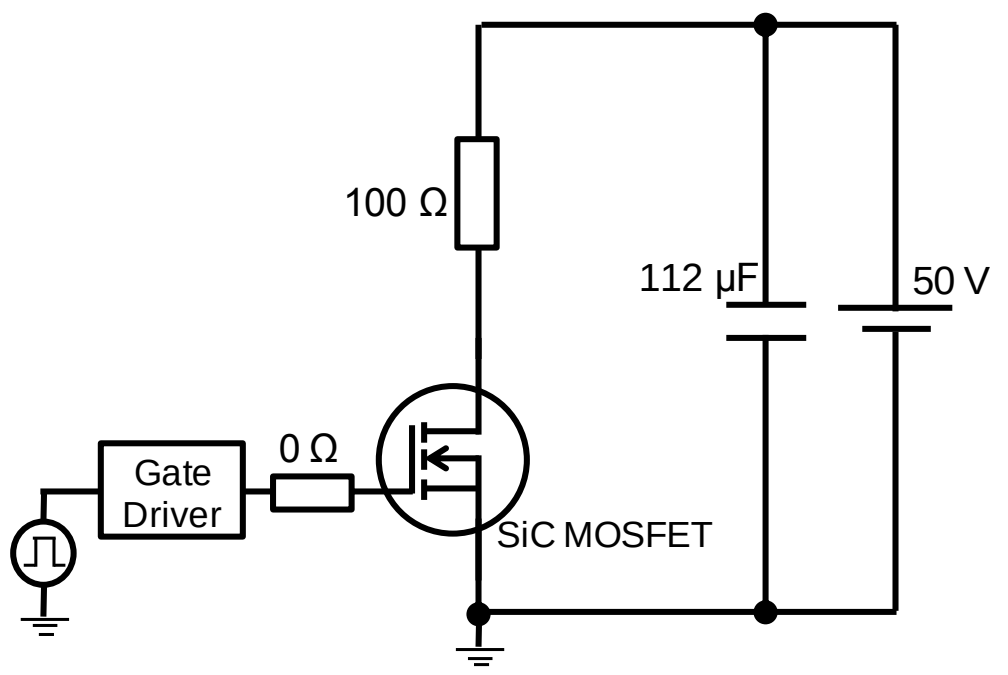

Fig.9 Schematic of a switching test circuit for SiC MOSFETs using our proposed gate driver.

The switching characteristics of the SiC MOSFETs driven by gate driver Prototype-B are shown in Fig. 10. The switching frequency was set at 20 MHz and the duty ratio of the control signal was 55% for both the high and low sides to suppress the penetration current through the GaN-HEMT push-pull configuration. We found that the gate voltage of the SiC MOSFET reached 18 V. Since on- and off-states of the SiC MOSFET were observed at 20 MHz, 20-MHz hard switching of SiC MOSFET was obtained using proposed gate driver Prototype-B. In our previous study [19], a gate driver based on Si suffered from thermal runaway at frequencies higher than 3 MHz. The gate driver with a GaN-HEMT push-pull configuration has higher driving capability and more robustness owing to the superior device performance of the GaN HEMTs. The switching waveforms of the SiC MOSFET exhibited slower turn-off characteristics than those of the turn-on. Those characteristics are mostly determined by the capacitance-voltage (C-V) characteristics of the SiC MOSFET. The limiting factors of the switching characteristics of the SiC MOSFET must be clarified to improve the device structure of the SiC MOSFET for enhancement of the switching frequency.

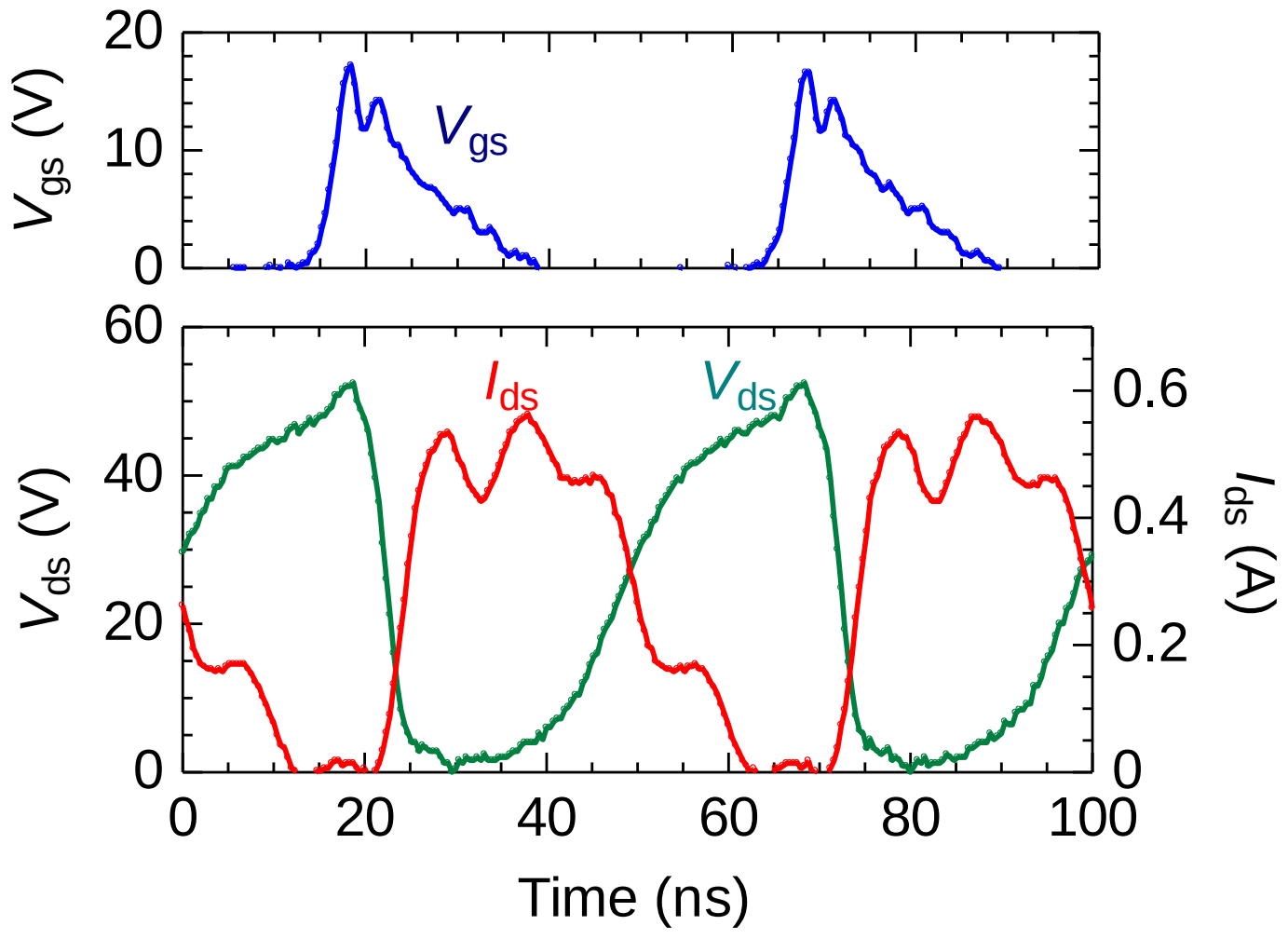

Fig.10 Output voltage ($V_{gGanL}$) of Si MOSFETs plotted against switching frequency. Duty ratio of input signal varied from 0.5 to 0.7. Supply voltage ($V_{dSiL}$) was 3.5 V.

V. CONCLUSION

We investigated the limiting factors of switching frequency in our proposed gate driver based on a GaN-HEMT push-pull configuration. The digital isolator did not work at high frequency, but complementary Si MOSFETs enhanced the driving capability, resulting in the 20-MHz hard switching of the SiC MOSFET. We successfully achieved the hard switching at 13.56 MHz (ISM band) using the proposed gate driver. In order to further improve the switching frequency, it is necessary to accelerate the digital isolator and reduce the power consumption of Si MOSFETs.

Acknowledgements

This research was partially supported by the Kyoto Super Cluster Program (JST) and the Cross-ministerial Strategic Innovation

Promotion Program (SIP, “Next-generation power electronics”(NEDO)). T. Okuda was supported by JSPS KAKENHI Grant Number JP16H06890.